\documentstyle[11pt,paspconf,epsf]{article}

\newcommand{\kms}{\mbox{${\rm km\ s^{-1}}$}}

\newcommand{\hbeta}{\mbox{${\rm H}\beta$}}

\newcommand{\z}{\mbox{${\rm [Z/H]}$}}
\newcommand{\feh}{\mbox{${\rm [Fe/H]}$}}

\newcommand{\enh}{\mbox{${\rm [E/Fe]}$}}
\newcommand{\afe}{\mbox{${\rm [\alpha/Fe]}$}}

\begin{document}

\title{The Ages of Early-Type Galaxies: A Cautionary Tale}

\author{S. C. Trager}
\affil{The Observatories of the Carnegie Institution of Washington,\\
813 Santa Barbara Street, Pasadena, CA 91101 USA}

\begin{abstract}
Early-type galaxies are not the simple Population II systems they have
long been assumed to be.  While upwards of 80\% of the stellar mass of
early-type galaxies likely formed at high redshift, small frostings of
intermediate-age stellar populations (a few to 20\% percent by mass of
1--2 Gyr old stars) are present in nearly every field and group
early-type galaxy and in at least some cluster early-types.  These
frostings of young stars have little effect on the determination of
photometric redshifts, thanks to the age-metallicity degeneracy of
broad-band colors, but even mild bursts of star formation at modest
redshifts (a few tenths) may make identification of the progenitors of
today's early-type galaxies difficult at cosmological distances.
\end{abstract}


\section{Introduction}

A driving force behind the present explosion in the use of photometric
redshifts is the need to ``preselect'' samples of interesting galaxies
(often at specific redshifts) from photometric catalogs for further
study.  Early-type galaxies have a pronounced photometric
signature---a strong Balmer break at 4000 \AA\ and red colors longward
of this---and thus are easily identified and their redshifts
determined from imaging surveys.  Studying the evolution of early-type
galaxies from relatively high redshift ($z\sim1$) to the present
should thus be rather simple, once a reasonably large sample of
galaxies at various redshifts are found.

This type of evolutionary study is predicated on a shaky assumption:
early-type galaxies are passively evolving, old stellar systems, an
assumption based on the suggestion by Baade (1963) that elliptical
(and by extension S0) galaxies are Population II systems.  Since the
early 1970's (e.g., Faber 1972, 1973; O'Connell 1980) it has been
clear that most elliptical galaxies have too much blue light to be
completely old populations.  However, age and metallicity are nearly
degenerate in the spectra of old ($>2$ Gyr) populations, and
sophisticated models and sensitive observations are required to break
this degeneracy.

In this talk I present evidence that the excess blue light in
early-type galaxies is due to small frostings of intermediate-age
($\sim1$ Gyr old) stars.  As these small frostings imply at least mild
starburst events at moderate redshifts ($z\sim0.1$--0.3), the
assumption of passively evolving early-type galaxies---especially in
the field and in groups---needs to be re-examined when interpreting
the results of photometric (and spectroscopic) redshift surveys.

\section{Breaking the Age-Metallicity Degeneracy}

\begin{figure}
\plottwo{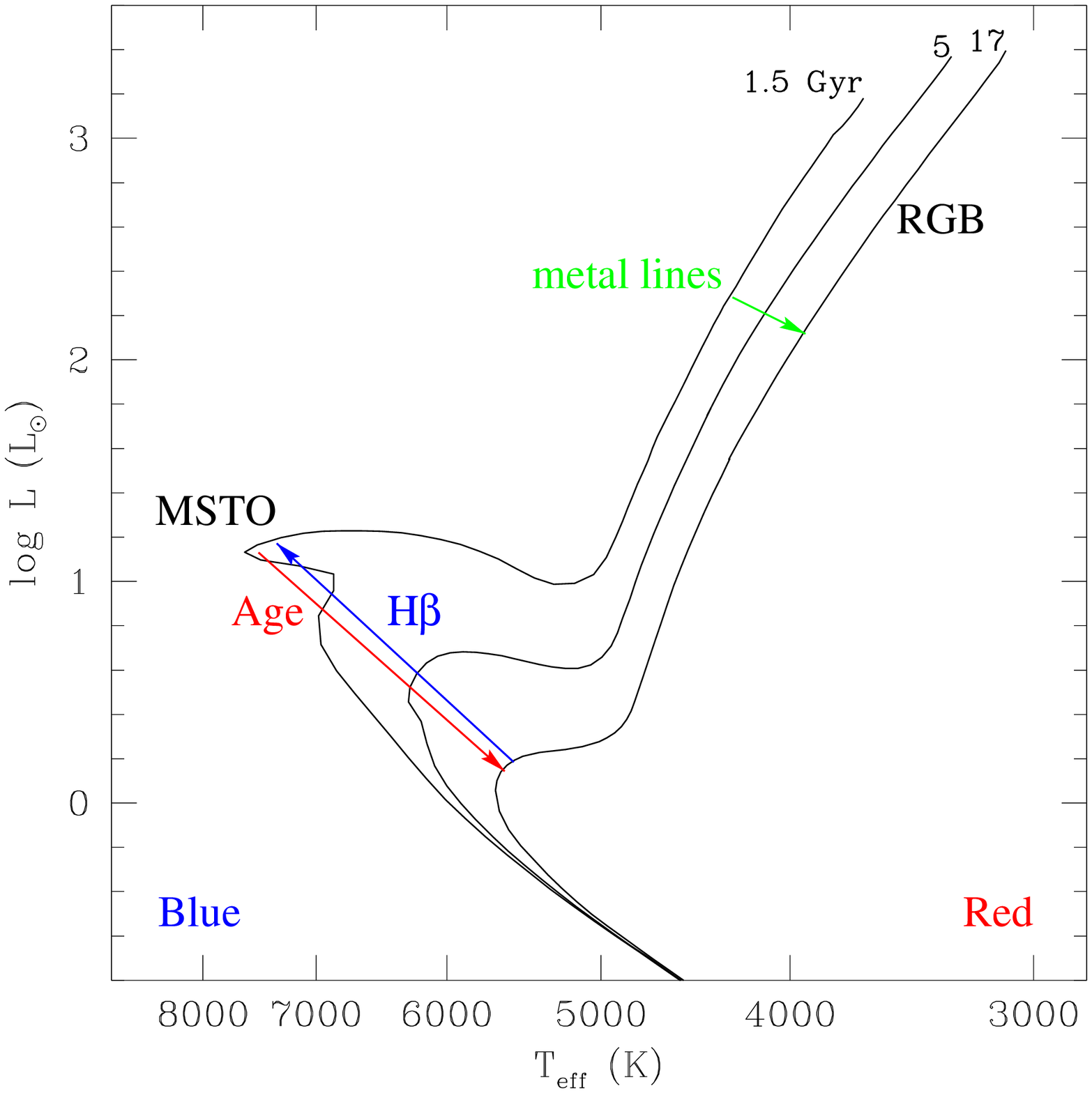}{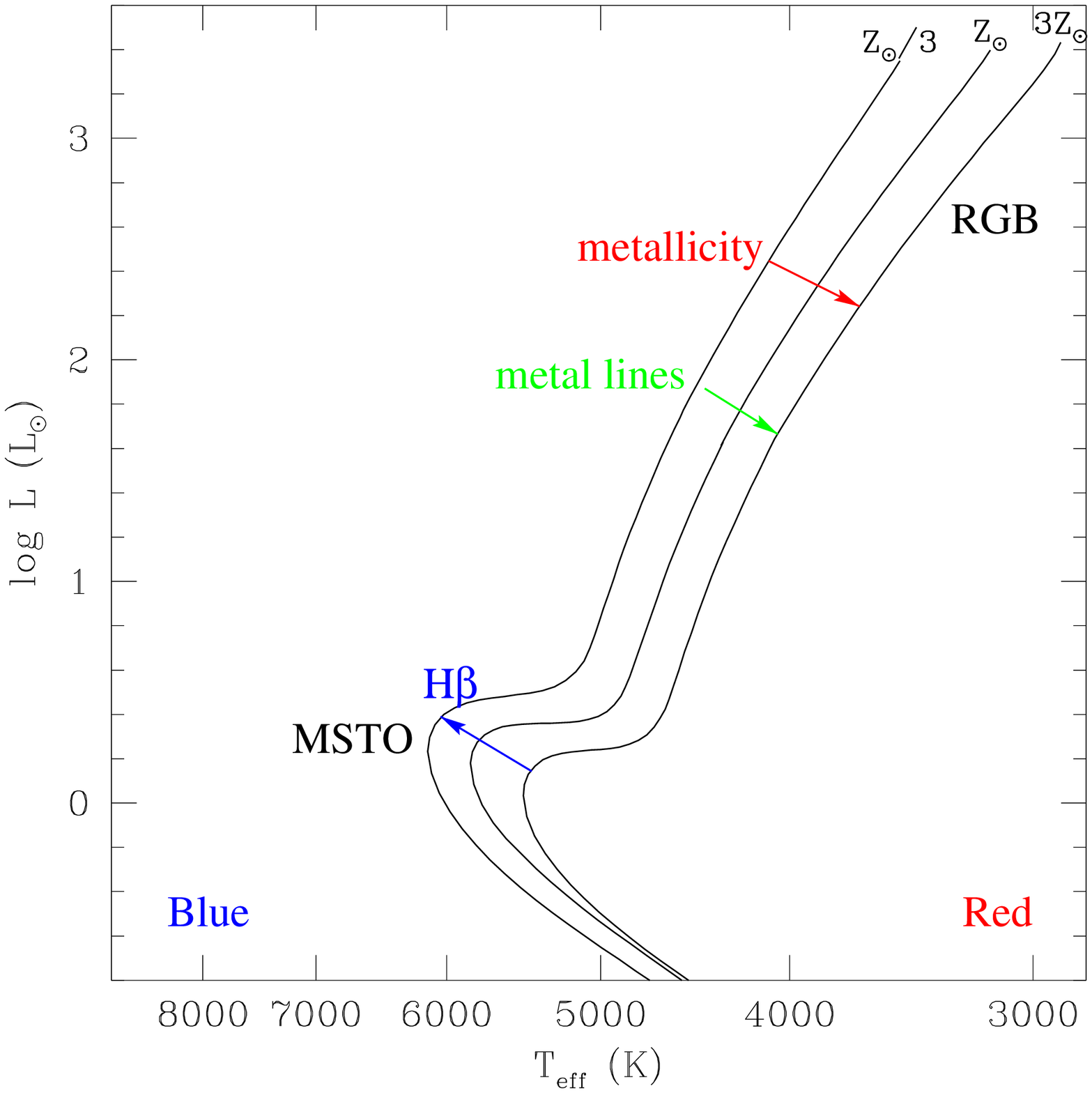}
\caption{The age-metallicity degeneracy and its resolution.  On the
left, solar metallicity isochrones of ages 1.5, 5, and 17 Gyr are
shown; on the right, 12 Gyr isochrones of metallicities $\z=-0.5$, 0.,
and $+0.5$ are shown (isochrones from Worthey 1994).  Broad-band
colors and metal-lines are dominated by the giant branches of these
populations, which respond nearly equally to age and metallicity
changes, and are therefore degenerate to compensating changes in these
parameters (to be precise, most broad-band colors and metal-line
strengths are constant when $\Delta\log Z/\Delta\log t=-3/2$).
Balmer-line strengths, like \hbeta, are dominated by light from the
main-sequence turnoff and therefore are much more sensitive to the age
of the population. By combining Balmer- {\it and\/} metal-line
strengths, we can break the age-metallicity degeneracy and determine
the ages of early-type galaxies
(Fig.~\ref{fig:local}).\label{fig:cmds}}
\end{figure}

As demonstrated in Figure~\ref{fig:cmds}, most broad-band colors
(particularly those longward of the 4000 \AA\ break) and metal-line
strengths are dominated by the light of the RGB of an old stellar
population.  Unfortunately, the RGB temperature (color) is degenerate
to compensating ages in age and metallicity.  This is the origin of
the notorious {\it age-metallicity degeneracy\/} (Faber 1972, 1973;
O'Connell 1980; Rose 1985; Renzini 1986).  This degeneracy is actually
quite helpful when determining photometric redshifts (Kodama, this
volume), but not when attempting to determine the evolutionary
histories of early-type galaxies.

However, Figure~\ref{fig:cmds} also demonstrates a way to break this
degeneracy: any color or absorption-line strength sensitive to the
temperature of the main sequence turn-off (MSTO) will be much less
sensitive to changes in metallicity than to changes in age.  In the
optical, the Balmer lines are the best tracers of the MSTO temperature
known.  Although there is still some residual dependence of the MSTO
temperature---and thus of the Balmer lines---on metallicity, Worthey
(1994) showed that the equivalent single-burst (``SSP-equivalent'')
age and metallicity of a stellar population can be read directly from
two-dimensional diagrams of Balmer- and metal-line strengths
(Fig.~\ref{fig:local}).

\section{The Stellar Populations of Early-Type Galaxies}

\begin{figure}
\plottwo{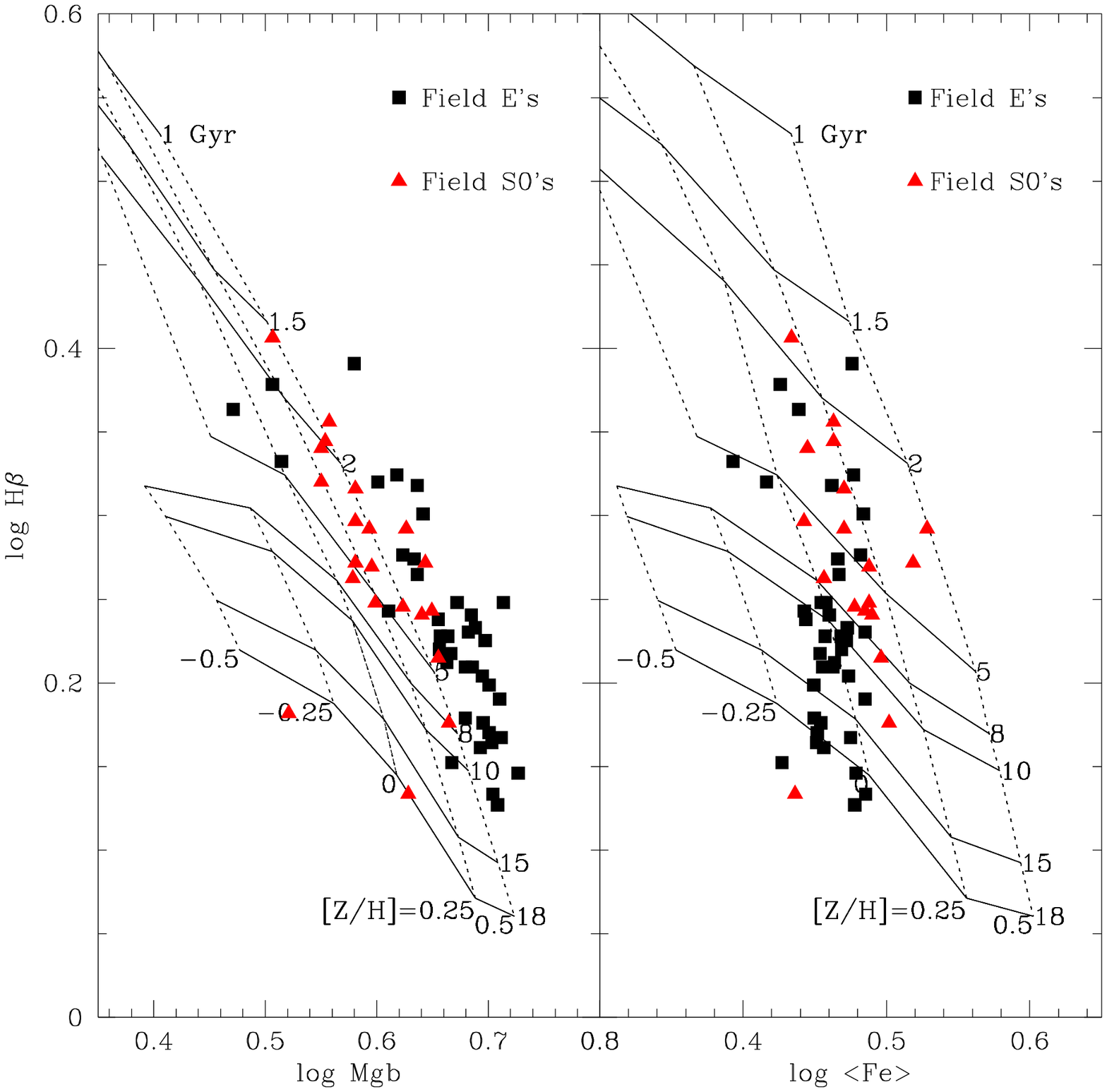}{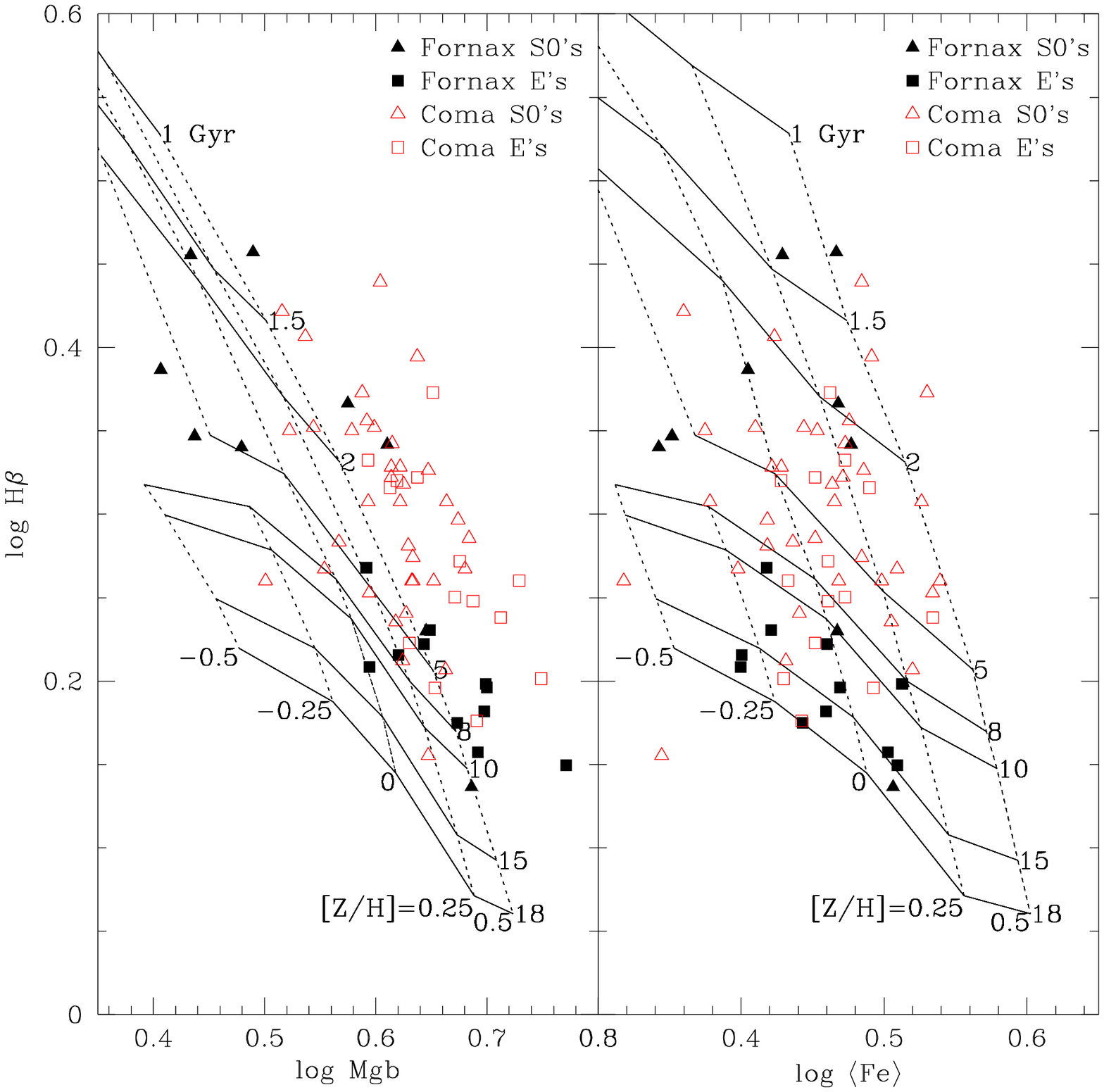}
\caption{Central absorption-line strengths of early-type galaxies.
(a) Field and group ellipticals (squares; Gonz\'alez 1993) and S0's
(triangles; Fisher, Franx \& Illingworth 1995).  (b) Cluster
early-types: Fornax (solid symbols; Kuntschner 1999) and Coma (open
symbols; J{\o}rgensen 1999).
\label{fig:local}}
\end{figure}

\begin{figure}
\plotone{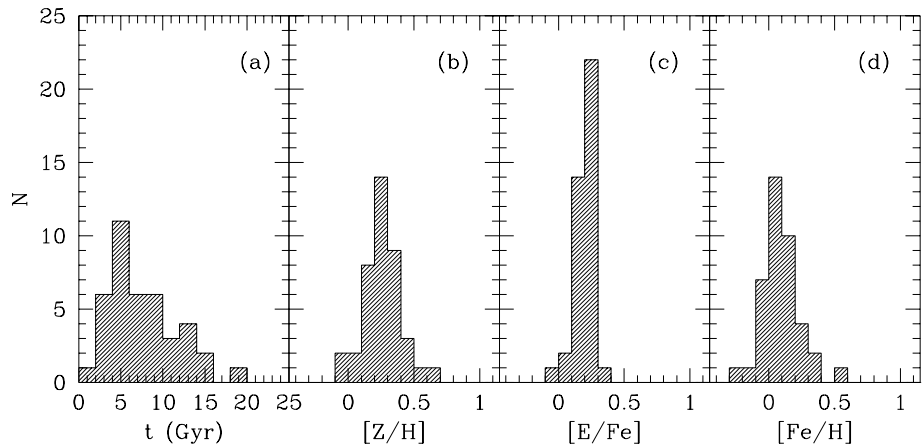}
\caption{The distribution of central stellar population parameters of
field and group elliptical galaxies (Trager et al.~1999). As expected
from Fig.~\ref{fig:local}, ellipticals span a large range in mean
stellar population age but narrow ranges of metallicity \z, iron
abundance \feh, and ``enhancement'' ratio \enh.\label{fig:hist}}
\end{figure}

Two-dimensional grids of Balmer- and metal-line strengths are shown in
Figure~\ref{fig:local} for local elliptical and S0 galaxies in the
field and in groups (Fig.~\ref{fig:local}a) and in the Fornax and Coma
clusters (Fig.~\ref{fig:local}b).  It is immediately obvious that
early-type galaxies are {\it not\/} uniformly old stellar populations
varying primarily in metallicity.  Even in dense clusters like Coma
and Fornax, many S0 and even a few elliptical galaxies have
SSP-equivalent ages of a few Gyr.  (This behavior has even been seen
in cluster populations at $z\sim0.4$--$0.8$; Trager 1997; Trager,
Faber \& Dressler, in prep.)  The distribution of stellar population
parameters (age, metallicity, enhancement ratio---akin to \afe---and
iron abundance) for a sample of 39 early-type galaxies (and the bulge
of M31) are shown in Figure~\ref{fig:hist} (Trager et al.~1999).

\begin{figure}
\plotone{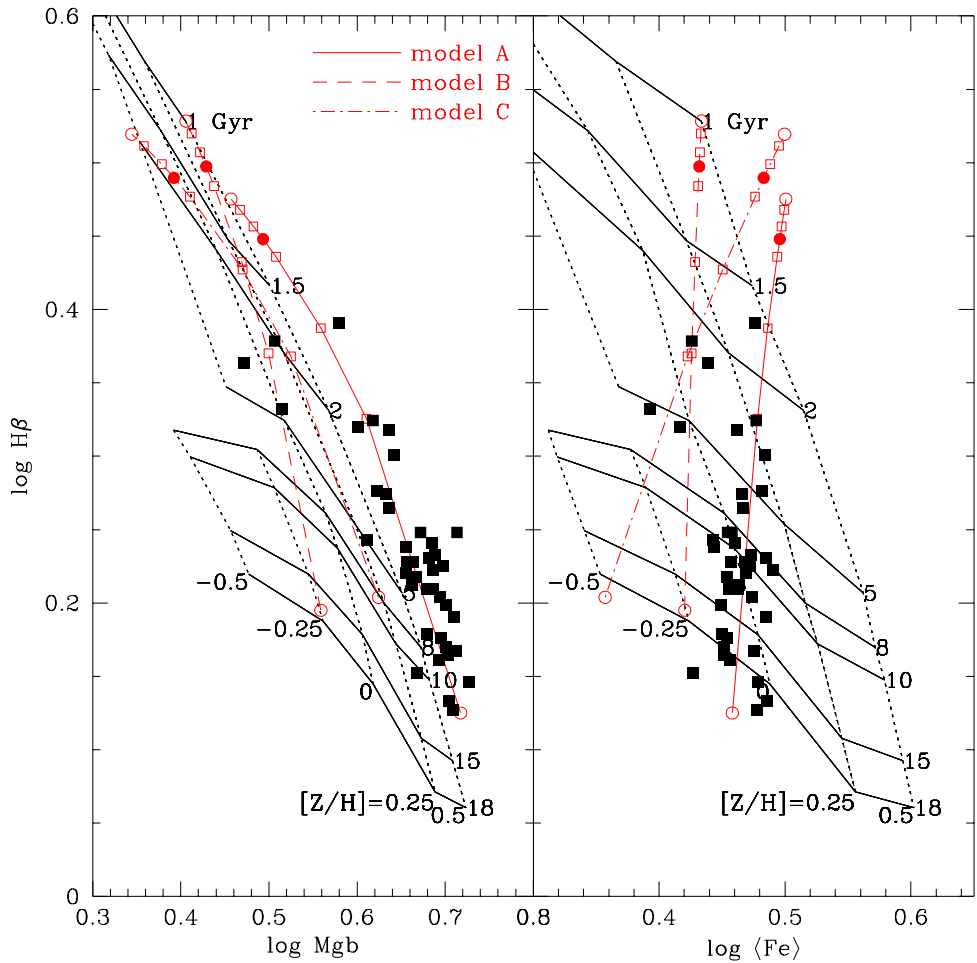}
\caption{The effects of multiple bursts on absorption-line strengths
(Trager et al.~1999).  Three models are shown, all consisting of
fractional amounts (by mass) of 1 Gyr burst population superimposed on
a 17 Gyr old progenitor population.  Model A is meant to represent a
plausible burst in a giant elliptical galaxy.  The other two models
are meant to represent plausible bursts in a small ($\sigma<150\kms$)
elliptical, one (model B) using solar abundance-ratio populations, the
other (model C) involving a metal-enriched wind ($\enh<0$) on top of a
SN II-rich progenitor ($\enh>0$).  Bursts of 10\%, 20\%, 40\%, 60\%,
and 80\% by mass are shown as open squares; bursts of 50\% by mass are
shown as solid circles.  Open squares are progenitor (lower) and burst
(upper) populations.  Solid squares are field ellipticals from
Gonz\'alez (1993), repeated from
Figure~\ref{fig:local}.\label{fig:twoburst}}
\end{figure}

These SSP-equivalent ages are difficult to interpret directly, as line
strengths of composite populations add like vectors weighted by the
light of each population.\footnote{Small amounts of blue straggler
stars and very metal-poor populations contributing blue horizontal
branch stars can also perturb the \hbeta\ strengths of early-type
galaxies but cannot seriously affect the \hbeta-strong galaxies seen
Figure~\ref{fig:local} (Trager et al.~1999).}
Figure~\ref{fig:twoburst} shows this effect graphically: even a small
burst of intermediate-age ($\sim1$ Gyr old) populations dramatically
increases the \hbeta\ strength of an old stellar population.  For
example, most elliptical galaxies in the field can be explained by a
small (a few to 10\% by mass) burst of a 1 Gyr population on top of a
very old, very massive stellar population; a 10\% burst by mass of a 1
Gyr old population on top of a 17 Gyr old population looks like a 2--3
Gyr old population (depending on the exact burst model chosen).  In
the most extreme cases (e.g., the field elliptical NGC 6702) the 1 Gyr
old burst required is as much as 20\% of the total stellar mass.

Note that burst strengths get much larger when the age of the burst
population is constrained to be much older than 1 Gyr due to the
rapidly increasing $M/L$ ratio of old stellar populations with age.
Burst ages much less than 1 Gyr are unlikely in most galaxies due to
the lack of significant amounts of hot star light seen at 4000 \AA\
(Rose 1985).  In multiple burst scenarios (two or more bursts on top
of an old population) the most recent burst, especially one that
occurred less than 1.5 Gyr ago, will dominate the final line
strengths.

\section{Conclusions and Cautions}

Local early-type galaxies in all environments, and even cluster S0
galaxies at redshifts out to at least $z\sim0.8$ (Trager, Faber \&
Dressler, in prep.), show evidence for at least small amounts of
recent star formation superimposed on old, massive progenitors.
Typically these frostings of intermediate-age stars comprise a few to
about 10\% of the mass of the total stellar population, but some
extreme cases may require up to 20\% or more of the stellar population
to have been formed 1--2 Gyr ago---i.e., at redshifts $z\la0.2$.

Although these bursts seem mild, even these mild starbursts will cause
significant spectrophotometric---and possibly morphological, if these
starbursts are dynamically induced---changes in the progenitors of
today's early-type galaxies at modest redshifts.  For a starburst
strength of 10\% by mass in an early-type galaxy, Charlot \& Silk
(1994) have demonstrated that the colors require roughly a Gyr to
revert to pre-burst levels (cf.~Schweizer \& Seitzer 1992).  However,
the morphological evolution may be much quicker than that---Mihos
(1995) has shown that a merger of two massive disk galaxies takes
about a Gyr to relax back into an early-type morphology; 10\%
accretion events are unlikely to take that long.  Because the amount
of mass required by the burst increases strongly with the present age
of the burst, these starbursts will become more dramatic with redshift
(if only a single secondary burst occurs in each galaxy, a simplistic
assumption).  In any case, the spectrophotometric and possible
morphological changes induced in even a mild burst will make
identification of at least some of the progenitors of present-day
elliptical galaxies a difficult task with photometric redshifts.  This
is just the ``progenitor bias'' (van Dokkum \& Franx 1996): if the
progenitors of some fraction of today's early-type galaxies were not
early-type galaxies at a given redshift, any sample of early-type
galaxies at that redshift would consequently be biased towards the
oldest galaxies.  The progenitor bias does not affect the detection
and identification of {\it old\/} early-type galaxies in photometric
redshift surveys, but it does make the interpretation of the evolution
of those galaxies more problematic.

\acknowledgements

I am grateful to my collaborators, Alan Dressler, Sandra Faber, Jesus
Gonz\'alez, and Guy Worthey for allowing me to present some of this
material in advance of publication.  I would also like to thank the
organizers, Robert Brunner, Marcin Sawicki, Lisa Storrie-Lombardi, and
Ray Weymann for an enjoyable (and convenient!) meeting.

\end{document}